\documentclass[conference]{IEEEtran}
\IEEEoverridecommandlockouts
\usepackage{cite}
\usepackage{amsmath,amssymb,amsfonts}
\usepackage{algorithmic}
\usepackage{graphicx}
\usepackage{textcomp}
\usepackage{xcolor}
\usepackage[a4paper, total={184mm,239mm}]{geometry}
\def\BibTeX{{\rm B\kern-.05em{\sc i\kern-.025em b}\kern-.08em
    T\kern-.1667em\lower.7ex\hbox{E}\kern-.125emX}}

\usepackage{tikz}

\usepackage{caption}
\usepackage{mathtools}

\usepackage{subcaption}

\usepackage{hyperref}
\usepackage{lipsum}

\newcommand\copyrighttext{%
  \footnotesize \textcopyright 2023 IEEE. Personal use of this material is permitted.
  Permission from IEEE must be obtained for all other uses, in any current or future
  media, including reprinting/republishing this material for advertising or promotional
  purposes, creating new collective works, for resale or redistribution to servers or
  lists, or reuse of any copyrighted component of this work in other works.
  }
  
\newcommand\copyrightnotice{%
\begin{tikzpicture}[remember picture,overlay]
\node[anchor=south,yshift=10pt] at (current page.south) {\fbox{\parbox{\dimexpr\textwidth-\fboxsep-\fboxrule\relax}{\copyrighttext}}};
\end{tikzpicture}%
}

\begin{document}

\title{A Mapping of Triangular Block Interleavers to DRAM for Optical Satellite Communication
}

%
\author{\IEEEauthorblockN{Lukas Steiner\IEEEauthorrefmark{2}, Timo Lehnigk-Emden\IEEEauthorrefmark{1}, Markus Fehrenz\IEEEauthorrefmark{1} and Norbert Wehn\IEEEauthorrefmark{2}}\\\vspace*{-10pt}
\IEEEauthorblockA{\IEEEauthorrefmark{1}\textit{Creonic GmbH}, Kaiserslautern, Germany\\
\{timo.lehnigk-emden, markus.fehrenz\}@creonic.com}
\IEEEauthorblockA{\IEEEauthorrefmark{2}\textit{Microelectronic Systems Design Research Group, University of Kaiserslautern-Landau}, Kaiserslautern, Germany\\
\{lukas.steiner, norbert.wehn\}@rptu.de}
}
%

\maketitle

\begin{abstract}
Communication in optical downlinks of low earth orbit (LEO) satellites requires interleaving to enable reliable data transmission.
These interleavers are orders of magnitude larger than conventional interleavers utilized for example in wireless communication.
Hence, the capacity of on-chip memories (SRAMs) is insufficient to store all symbols and external memories (DRAMs) must be used. 
Due to the overall requirement for very high data rates beyond 100\,Gbit/s, DRAM bandwidth then quickly becomes a critical bottleneck of the communication system.
In this paper, we investigate triangular block interleavers for the aforementioned application and show that the standard mapping of symbols used for SRAMs results in low bandwidth utilization for DRAMs, in some cases below 50\,\%.
As a solution, we present a novel mapping approach that combines different optimizations and achieves over 90\,\% bandwidth utilization in all tested configurations.
Further, the mapping can be applied to any JEDEC-compliant DRAM device.
\end{abstract}

\copyrightnotice

\begin{IEEEkeywords}
Optical communication, Interleaving, DRAM
\end{IEEEkeywords}

\section{Introduction and Related Work}
Interleaving is a key method in communication systems to correct burst errors that occur during transmission.
In free-space optical communication from \textit{low earth orbit} (LEO) satellites to earth, huge amounts of data have to be transmitted via lasers in a short time frame. 
This leads to data rate requirements beyond 100\,Gbit/s which have to be decoded in real time.
Due to high dynamics in channel quality and a long coherence time of the optical communication channel ($>$ 2\,ms), large interleavers are necessary to enable reliable transmission at target code rates. 
A suitable interleaver type for this application is the \textit{triangular block interleaver} where the symbols of multiple consecutive code words are first \textit{written} to a triangular (upper left half of a square) storage array in \textit{row-wise} order and afterwards \textit{read} from the array in \textit{column-wise} order.
State-of-the-art interleavers are implemented by corresponding memory addressing, i.e., row-wise writing to an SRAM and column-wise reading.
Unlike SRAM, however, the achievable transfer bandwidth of DRAM depends strongly on the access pattern. 
Patterns with few address bit toggles (e.g., sequential or strided) enable high bandwidth utilization, while patterns with many toggles (e.g., random) decrease bandwidth utilization significantly due to frequent page misses.
If the two-dimensional index space of a triangular block interleaver is mapped to linear DRAM storage in row-major order as in the case of an SRAM implementation, the write phase will result in a sequential pattern (high bandwidth utilization), but the read phase will result in frequent page misses (low bandwidth utilization).
Since the maximum throughput of the interleaver is not determined by the average, but by the minimum DRAM bandwidth utilization across both access phases, the theoretical maximum bandwidth of a DRAM configuration must be largely oversized (faster speed grade or wider data bus) to match the requirements of the communication system.
This leads to higher costs and additional energy consumption.

In previous research, several optimized mappings of two-dimensional index spaces to DRAM for alternate row- and column-wise accesses have been proposed~\cite{siche12,zhodou10,garpir20,akifra15,kimpar00,akimil12,lanpir12,pra15,sinpra15}. 
However, none of the works considered the impact of bank groups, which have been introduced in all recent JEDEC DRAM standards (e.g., DDR4/5, LPDDR5, HBM).

In this work, we present a novel general approach to map the two-dimensional index space of triangular block interleavers to any JEDEC-compliant DRAM device.
By combining three optimizations, we eliminate the major limiting factors for DRAM bandwidth utilization in both access directions.
We prove the effectiveness of our mapping with cycle-accurate simulations of five different DRAM standards and two different speed grades per standard.
\section{Optimized Mapping}
\begin{figure}
  \begin{subfigure}[b]{0.49\linewidth}
  \centering
    \includegraphics[width=.51\linewidth]{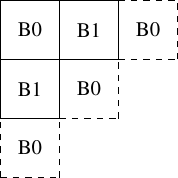} 
    \caption{Banks}
    \label{fig:bank}
  \end{subfigure}
  \begin{subfigure}[b]{0.49\linewidth}
  \centering
    \includegraphics[width=.34\linewidth]{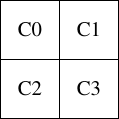} 
    \caption{Columns}
    \label{fig:column}
  \end{subfigure}
  \par\bigskip%
  \begin{subfigure}[b]{0.49\linewidth}
  \centering
    \includegraphics[width=.85\linewidth]{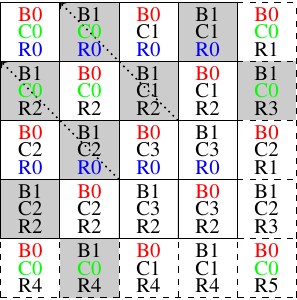} 
    \caption{Banks, Columns and Rows}
    \label{fig:rcb1}
  \end{subfigure}
  \begin{subfigure}[b]{0.49\linewidth}
  \centering
    \includegraphics[width=.85\linewidth]{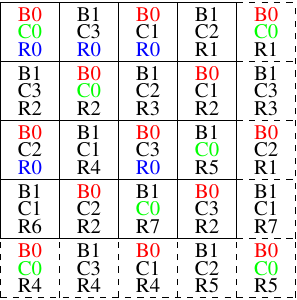} 
    \caption{BCR with Column Offset}
    \label{fig:rcb2}
  \end{subfigure}
\caption{Optimized Mapping Schemes}
\label{fig:figures}
\end{figure}
It is important to note that the number of bits accessed with a single DRAM burst is much larger than one symbol (e.g., 512 bits vs. 3 bits).
In our application, interleaving is therefore divided into two stages, and a small SRAM block interleaver first ensures that symbols within a DRAM burst belong to different code words.
The optimized mapping assigns each position of the two-dimensional index space of the interleaver to a DRAM address composed of a bank, column and row.
In order to handle standards both with and without bank groups using the same mapping, it is presumed that the lower bank address bits always denote the bank group.

For a DRAM standard with bank groups, the full bandwidth can only be utilized if the \textbf{bank group is switched with each access} in a round-robin order.
This is achieved by incrementing the bank index by one with each access in both directions, which leads to the diagonal pattern shown in Fig.~\ref{fig:bank} exemplarily for two banks.

Irrespective of the DRAM standard, \textbf{page misses should be minimized} for high bandwidth utilization.
To divide them equally between both access directions, the two-dimensional index space is partitioned into small rectangles, each of which is assigned to a DRAM page.
This mapping is shown in Fig.~\ref{fig:column} for a page with four columns.

Combining both optimizations and adding DRAM rows leads to the mapping shown in Fig.~\ref{fig:rcb1}.
For reasons of space, only the top left quarter of the index space is presented.
Bank 0, column 0 and row 0 are colored for a better understanding.
One final problem is that the remaining page misses occur almost simultaneously on all banks, which prevents a page miss on one bank from being effectively masked by page hits on other banks.
As a solution, \textbf{page misses on different banks are staggered} by introducing a bank-dependent column offset.
Simply put, all positions of the index space are shifted to the top left by an offset that depends on the bank they are mapped to.
In Fig.~\ref{fig:rcb1}, all positions mapped to bank 1 are shifted by two (offset of one column in both directions, dotted arrows), while the offset is zero for positions mapped to bank 0.
In this way, the first page miss on bank 1 occurs one access earlier than on bank 0.
The shifting is done circular, i.e., positions with a gray background are moved to the bottom and right border, respectively.
This leads to the final mapping shown in Fig.~\ref{fig:rcb2}.\footnote{Figs.~\ref{fig:rcb1} and \ref{fig:rcb2} show a rectangular index space, which would result in a storage overhead of more than 50\,\% for a triangular block interleaver. In reality, the number of DRAM rows decreases from the top to the bottom of the index space, which allows a more storage-efficient mapping.}
Due to space limitations, the mapping rules are not elaborated further.
However, they only consist of additions, logical shifts and bitwise operations, which enables a hardware implementation with low complexity.
\section{Experimental Results} 
\begin{table}
\caption{DRAM Bandwidth Utilizations}
\label{tab:bw}
\centering
\resizebox{.9\linewidth}{!}
{
\begin{tabular}{|c|c|c|c|c|}
\hline
\textbf{DRAM} & \multicolumn{2}{|c|}{\textbf{Row-Major Mapping}} & \multicolumn{2}{|c|}{\textbf{Optimized Mapping}}\\
\cline{2-5} 
\textbf{Configuration} & \textbf{Write} & \textbf{Read} & \textbf{Write} & \textbf{Read} \\
\hline
  DDR3-800  & \textbf{95.99\,\%} & 96.03\,\% & \textbf{95.99\,\%} & 96.26\,\% \\
  DDR3-1600 & 95.75\,\% & \textbf{64.16\,\%} & \textbf{95.91\,\%} & 96.16\,\% \\
  DDR4-1600 & 92.02\,\% & \textbf{73.92\,\%} & \textbf{92.01\,\%} & 92.37\,\% \\
  DDR4-3200 & 91.83\,\% & \textbf{43.50\,\%} & \textbf{91.86\,\%} & 92.15\,\% \\
  DDR5-3200 & 100.00\,\% & \textbf{96.37\,\%} & \textbf{100.00\,\%} & 100.00\,\% \\
  DDR5-6400 & 99.90\,\% & \textbf{88.95\,\%} & \textbf{99.83\,\%} & 99.97\,\% \\
LPDDR4-2133 & 99.02\,\% & \textbf{66.00\,\%} & 99.41\,\% & \textbf{98.30\,\%} \\
LPDDR4-4266 & 98.03\,\% & \textbf{35.77\,\%} & \textbf{99.67\,\%} & 99.72\,\% \\
LPDDR5-4267 & 99.39\,\% & \textbf{55.87\,\%} & \textbf{99.77\,\%} & 100.00\,\% \\
LPDDR5-8533 & 97.56\,\% & \textbf{47.25\,\%} & \textbf{99.14\,\%} & 99.66\,\% \\
\hline
\end{tabular}
}
\end{table}
To quantify the improvements, multiple configurations are simulated with the cycle-accurate DRAM simulator \mbox{DRAMSys}~\cite{stejun20}.
Table~\ref{tab:bw} shows the achieved bandwidth utilizations for a triangular block interleaver with 12.5\,M elements.
Results for other interleaver dimensions are omitted for space reasons because they differ only slightly.
The bandwidth utilization is separated into write (row-wise) and read (column-wise) phase since the minimum of both (bold) determines the maximum throughput of the interleaver.
Small variations within each phase are balanced by buffers in the memory controller.

The optimized mapping achieves significant improvements in bandwidth utilization over the row-major mapping for the majority of configurations, in some cases from below 50\,\% to over 90\,\%.
Especially the faster speed grade device of each standard and LPDDR devices benefit greatly.
The missing percentages of the optimized mapping are caused by refresh commands.
When refresh is disabled, which is legal as long as the maximum lifetime of the interleaver data is smaller than the DRAM refresh period (between 32\,ms and 64\,ms), a bandwidth utilization of over 99\,\% is consistently achieved.
The proposed optimizations can also be applied to different applications with similar DRAM access patterns.
%

%
\end{document}